\documentclass[12pt]{iopart}
\usepackage{iopams}

\newcommand{\rmr}{\mathrm{Re}}
\newcommand{\dzp}{\frac{\partial P}{\partial z}}
\newcommand{\q}{\bPsi}

\begin{document}

\title{Statistics of a noise-driven Manakov soliton}

\author{Stanislav A Derevyanko$^{1,2}$, Jaroslaw E Prilepsky$^{3}$, and Dennis A Yakushev$^{2}$}

\address{$^1$Photonics Research Group, Aston University,
Aston Triangle, B4 7ET, Birmingham, United Kingdom}

\address{$^2$Institute for Radiophysics and Electronics National
Academy of Sciences of Ukraine,\\ 12 Acad. Proscuri St., 61085,
Khrakov, Ukraine}

\address{$^3$B.I. Verkin Institute for Low Temperature Physics and
Engineering, National Academy of
Sciences of Ukraine, 47 Lenin Ave., 61103, Kharkov, Ukraine}%

\eads{\mailto{s.derevyanko@aston.ac.uk},
\mailto{astrayed@yandex.ru}, \mailto{den@ire.kharkov.ua}}

\date{\today}

\begin{abstract}
We investigate the statistics of a vector Manakov soliton in the
presence of additive Gaussian white noise. The adiabatic
perturbation theory for Manakov soliton yields a stochastic
Langevin system which we analyze via the corresponding
Fokker-Planck equation for the probability density function (PDF)
for the soliton parameters. We obtain marginal PDFs for the
soliton frequency and amplitude as well as soliton amplitude and
polarization angle. We also derive formulae for the variances of
all soliton parameters and analyze their dependence on the initial
values of polarization angle and phase.
\end{abstract}

\submitto{\JPA}

\maketitle

\section{\label{sec:intro} Introduction}

Nonlinear Shrodinger equation (NLSE) and its vector modifications
have been studied in various contexts for more than 30 years. It
has found applications in such areas as plasma physics,
hydrodynamics, magnetism and nonlinear fibre optics. The NLSE, the
so-called modified NLSE, and some of their vector counterparts
(e.g. Manakov equations) are integrable and, as such, possess
stable soliton solutions that exist due to the precise balance
between the effects of nonlinearity and dispersion. Since solitons
preserve their shape and internal properties during the
propagation they have been widely used in fibre optics as
information carriers \cite{Agraval,Hasegawa}.

In most applications, however, integrable systems represent only
the leading approximation of the real physical model. Quite often
such models can be analysed using the concept of a \textit{nearly
integrable system}. In such systems the deviations from the
initial integrable model are considered to be small and the
underlying dynamics can be inferred via methods of perturbation
theory \cite{YuraBoris}. The important and distinct class of
perturbations constitute random perturbations of different origin.
In particular, in fibre optics, stochastic perturbations emulate
the effects of the amplifier spontaneous emission as well as
various fibre inhomogeneities \cite{Mecozzi,WaiMenyuk,LakobaKaup}.
The influence of random perturbations on solitons has been studied
in different integrable systems including Korteveg-de-Vries (KdV)
equation \cite{KdV1,KdV2,KdV3}, NLSE
\cite{GH,Magomet,Falk,OL,JOSAB} (see the latter Ref. for
additional references), modified NLSE \cite{doktor-EL, doktorov1,
Yakushev}, and Manakov system \cite{WaiMenyuk, LakobaKaup,
doktorov1}. Inasmuch as solitons in integrable systems are robust
against small perturbations, the weak noise leads to random walk
of soliton parameters (which is called jitter) as well as
generation of dispersive radiation. In the sequel we will only be
concerned with the additive white Gaussian noise (AWGN) model,
that represents the simplest physically justified example of
random perturbation. For such systems most celebrated result for
soliton jitters is the so-called Gordon-Haus effect: the random
walk of the soliton position due to the effect of weak AWGN
\cite{GH}. The main tool used for studying the statistics of the
soliton parameters for the aforementioned systems is the adiabatic
soliton perturbation theory (see e.g. review paper
\cite{YuraBoris} and references therein). Since the perturbation
has stochastic nature the resulting system of equations for
soliton parameters represents a \textit{Langevin system of
stochastic equations} with the multiplicative noise. For this
Langevin system, in the vast majority of works on the subject the
noise was assumed to be additive, and the equations were
linearised. The assumption of the additive noise in the linearised
Langevin equations can give rise only to the \textit{Gaussian
probability dencity functions} (PDF). However the exact statistics
of the NLSE soliton, for example, in the presence of AWGN are not
Gaussian as was shown in papers \cite{Magomet,Falk,OL,JOSAB}.
Gaussian approximation can only be used for approximate estimates
of the integral characteristics of the PDF, like e.g. variances of
soliton jitters.

In the present paper we will consider an integrable Manakov system
driven by vector AWGN, a system which, in some respect, is a
vector generalisation of the NLSE. Soliton of the Manakov
equations possesses two additional parameters as compared to NLSE
soliton. Doktorov and Kuten \cite{doktorov1} demonstrated in the
Gaussain approximation that timing jitter of the Manakov soliton
can be managed by adjusting these two extra parameters. In the
current paper we will show that not only timing jitter but all the
remaining jitters can be managed in the similar fashion provided
that the vector AWGN has anisotropic statistical properties. We
will also obtain (for the first time to our knowledge) marginal
PDFs for the parameters of the Manakov soliton, that describe
correctly the \textit{non-Gaussian} statistics of rare
fluctuations. We will derive the exact statistics of the soliton
parameters using the combination of the adiabatic soliton
perturbation theory and the Fokker-Planck equation (FPE) approach,
proceeding in the same way as in Refs.\cite{OL,JOSAB} for the NLSE
soliton.

The paper is organized as follows: in Section \ref{sec:pertub} we
derive a system of Langevin equations with multiplicative noise
using well established technique of the adiabatic perturbation
theory for a single Manakov soliton. In section \ref{sec:marginal}
the Fokker-Planck equation attached to the corresponding Langevin
system is derived. In the same section we find the exact solution
for the PDF of the amplitude-frequency jitter. Then we use a WKB
approximation for the FPE to obtain a marginal PDF for the soliton
amplitude and polarization angle. Eventually, in Section
\ref{sec:Gaussian} the Gaussian limit is considered -- we
calculate variances for all the soliton parameters. In Conclusion
we summarise the results obtained.

\section{Stochastic adiabatic perturbation theory for Manakov
soliton}%
\label{sec:pertub}

Consider a vector soliton propagating along a random medium, which
is described by the perturbed Manakov equations
\begin{equation}
\rmi \bPsi_{z}+\bPsi_{tt}+2\,(\bPsi\bPsi^{\,*})\,\bPsi=\mathbf{n}
 \label{man-eq} \, ,
\end{equation}
where the components of vector $\bPsi = (u(z,t),v(z,t))^T$
represent two polarization states of a propagating vector field
(e.g. electrical field in a birefringent optical fibre), whereas
the r.h.s. of Eq.(\ref{man-eq}) accounts for the random noise
(e.g. ASE) and its components have the following correlation
properties:
\begin{eqnarray}
 \langle n_{\alpha}(t,z) \rangle = \langle n_{\alpha}(t,z)n_{\beta}(t',z')
\rangle = 0 \, ,
\\ \label{corrmatrix}
 \langle n_{\alpha}(t,z)n^{*}_{\beta}(t',z') \rangle =
D_{\alpha\beta}\delta(z-z')\delta(t-t') \,, \qquad
\alpha,\beta=1,2.
\end{eqnarray}
Here matrix $\hat{D}$ is symmetric and positive definite.

In the absence of any perturbation the Manakov equations belong to
the class of those which are integrable by the inverse scattering
transform method. The general form of the one-soliton solution of
Manakov equations $\bPsi_0$ is given by following expressions (see
e.g. \cite{LakobaKaup}):
\begin{eqnarray}
\fl u(z,t) =
 2\eta(z)\cos\beta(z)  \mathrm{sech}\{2\eta (z)\left[ t  - T(z) \right] \}
 \exp \{ \rmi \alpha(z) - 2\rmi\left[ t  - T(z) \right] \xi(z) + \rmi \phi(z) \} ,
\label{u}
\\
\fl v(z,t)  = 2\eta (z)\sin\beta (z) \mathrm{sech}\{2\eta
(z)\left[ t  - T(z) \right] \}\exp\{\textrm{i}\alpha (z) -
2\rmi\left[ t  - T(z) \right] \xi (z) - \rmi \phi(z)\} , \label{v}
\end{eqnarray}
and dependencies of the six soliton parameters on the propagation
distance $z$ are: \numparts
\begin{eqnarray}
\eta(z) = \eta_{0},\\
\xi(z) = \xi_{0},\\
\alpha(z) = 4(\eta_{0}^{2}+\xi_{0}^{2} )z + \alpha_{0},\\
T(z) = -4\xi_{0} z + T_{0},\\
\varphi(z) = \varphi_{0},\\
\beta(z) = \beta_{0},
\end{eqnarray}
\endnumparts
where the constants with zero subscripts are the values of soliton
parameters at $z=0$.

Because the noise intensity is always small, the noise term,
$\mathbf{n}(t,z)$, can be considered as a perturbation. In the
presence of an arbitrary perturbation,
$\mathbf{R}=(R_{1},R_{2})^T$, the soliton parameters become slowly
varying functions of propagation distance $z$ and their evolution
during propagation should be determined via the equations of the
adiabatic soliton perturbation theory
\cite{LakobaKaup,midrio,doktorov-2}: \numparts
\begin{eqnarray}
\label{eta}
 \frac{\rmd\eta}{\rmd z}&=&\rmr\int\rmd
  t\,\mathbf{g}_{\eta}^{\,*}(t,z)\mathbf{R}(t,z),\\
\frac{\rmd\xi}{\rmd z} &=&\rmr\int\rmd t\,
 \mathbf{g}_{\xi}^{\,*}(t,z)\mathbf{R}(t,z),\\
\frac{\rmd\alpha}{\rmd z}&=&4(\eta^{2}+\xi^{2})+\rmr\int\rmd t\,
 \mathbf{g}_{\alpha}^{\,*}(t,z)\mathbf{R}(t,z),\\
\frac{\rmd T}{\rmd z}&=&-4\xi+\rmr\int\rmd t\,
 \mathbf{g}_{T}^{\,*}(t,z)\mathbf{R}(t,z),\\
\frac{\rmd\varphi}{\rmd z}&=&\rmr\int\rmd t\,
 \mathbf{g}_{\varphi}^{\,*}(t,z)\mathbf{R}(t,z),\\
\frac{\rmd\beta}{\rmd z}&=&\rmr\int\rmd t\,
 \mathbf{g}_{\beta}^{\,*}(t,z)\mathbf{R}(t,z) \label{beta},
\end{eqnarray}
\endnumparts
where vector functions $\mathbf{g}_i$ can be expressed through
derivatives of the single-soliton ansatz $\bPsi_0$ with respect to
appropriate soliton parameters. These functions can be written in
the compact form as follows:
\begin{eqnarray} \nonumber
\mathbf{g}_{\eta}=\frac{1}{2}\frac{\partial\q_0}{\partial\alpha}
\, , \qquad \mathbf{g}_{\xi}=-\frac{\xi}{\eta}\,\mathbf{g}_{\eta}+
\frac{1}{4\eta}\frac{\partial\q_0}{\partial T} \, , \\
\nonumber
\mathbf{g}_{\varphi}=\frac{\csc(2\beta)}{4\eta}\frac{\partial\q_0}{\partial\beta}
\,  , \qquad
\mathbf{g}_{\alpha}=-\cos(2\beta)\mathbf{g}_{\varphi}+
\frac{\xi}{2\eta}\frac{\partial\q_0}{\partial\xi}-
\frac{1}{2}\frac{\partial\q_0}{\partial\eta} \, ,
\\ \nonumber
\mathbf{g}_{T}=-\frac{1}{4\eta}\frac{\partial\q_0}{\partial\xi} \,
, \qquad \mathbf{g}_{\beta}=\frac{\cot(2\beta)}{2\eta}
\mathbf{g}_{\eta}-
\frac{\csc(2\beta)}{4\eta}\frac{\partial\bPsi_0}{\partial\varphi}
\, .
\end{eqnarray}
The domain of integration in (\ref{eta})-(\ref{beta}) should be
from $-\infty$ to $\infty$. The set of equations
(\ref{eta})-(\ref{beta}) with $\mathbf{R}(t,z)=\mathbf{n}(t,z)$
describes the fluctuations of soliton parameters under the action
of AWGN provided that the interaction between the soliton and
background radiation is negligible. The perturbation theory is not
applicable in the limit $\eta\rightarrow 0$ and also in the limit
when the polarization angle $\beta$ approaches values $\pi n/2$
with an integer $n$.

\section{Marginal PDFs for soliton parameters}
\label{sec:marginal}%
In this section we proceed to derive the FPE for all six soliton
parameters and obtain the exact statistics of the
amplitude-frequency jitter. We also use a WKB approach to derive
marginal PDF for soliton amplitude and polarization angle. For
simplicity we consider the case where the components of
perturbation vector $\mathbf{n}$ are two independent noises with
the same intensity $D$:
\begin{equation}
 \langle n_{\alpha}(t,z)n^{*}_{\beta}(t',z') \rangle =
  D\delta_{\alpha \beta}\delta(z-z')\delta(t-t') \, , \qquad
  \alpha, \beta = 1, 2 \, .
\end{equation}
In that case the FPE reads as (see Appendix):
\begin{eqnarray} \fl \nonumber
\frac{\partial P}{\partial z} =
   &\frac{\partial^2}{\partial\eta^2} \left[\frac{D\,\eta}{4}P\right]+
    \frac{\partial^2}{\partial\xi^2}\left[\frac{D\,\eta}{12}P\right]+
     \frac{\partial^2}{\partial T^2}\left[\frac{D\,{\pi }^2}{192\,{\eta}^3}P\right]
+\frac{\partial^2}{\partial \varphi^2}\left[\frac{D\,{\csc
(2\,\beta )}^2}{16\,\eta
      }P\right]+
\\ \nonumber \fl
&\frac{\partial^2}{\partial \beta^2}\left[\frac{D}{16\,\eta
         }P\right]
+\frac{\partial^2}{\partial \alpha^2}\left[
   \frac{D}{4}\,\left(\frac{1}{3\,\eta }
     +\frac{\pi^2(\eta^2+3\xi^2)}{36\eta^3}+
    \frac{\cot^2 (2\,\beta )}{4\,\eta}
\right)P\right]-
\\ \nonumber \fl
&\frac{\partial^2}{\partial \varphi \partial \alpha} \left[\frac{D
\cot (2\beta) \csc(2\beta)}{16 \eta}P\right]
-\frac{\partial^2}{\partial T \partial \alpha} \left[\frac{D \pi^2
\xi}{96 \eta^3}P\right]
-\frac{\partial}{\partial\eta}\left[\frac{3D}{4}P\right]+\frac{\partial}{\partial
T} \left[4\xi P \right]-
\\ \fl
&\frac{\partial}{\partial\beta}\left[\frac{D\cot(2\beta)}{8\eta}P\right]-
         \frac{\partial}{\partial \alpha} \left[ 4(\eta^2+\xi^2)
         P\right].
          \label{FPE}
\end{eqnarray}
The boundary conditions in angular variables $\alpha,\varphi$ and
$\beta$ are those of periodicity. In $\xi$, $\eta$ and $T$ the PDF
must decrease rapidly enough to provide normalization. At $\eta=0$
we assume that the $\eta$ component of the probability density
current vanishes (see below) which means that no solitons are
created of annihilated.

\subsection{Marginal PDF for amplitude-frequency jitter}
One can obtain in a closed form an autonomous FPE for the PDF of
the amplitude-frequency jitter $P(\eta,\xi| \, z)$. Thin can be
done by integrating out the redundant degrees of freedom in
(\ref{FPE}) by virtue of boundary conditions. Then the FPE for the
PDF of the amplitude-frequency jitter is
\begin{equation}
 \dzp=
 -\frac{\partial}{\partial\eta}\left[\frac{3D}{4}P\right]+
  \frac{\partial^2}{\partial\eta^2}\left[\frac{D\eta}{4}P\right]+
  \frac{\partial^2}{\partial\xi^2}\left[\frac{D\eta}{12}P\right].
\end{equation}
Since, without loss of generality, the initial value of $\xi$ may
be put equal to zero the latter equation must be solved under the
initial condition
\begin{equation}
 P(\eta,\xi| \, 0)=\delta(\xi)\delta(\eta-\eta_{0}).
\end{equation}
The $\eta$-component of the probability current vector is zero in
the plane $\eta=0$:
\begin{equation}
 \left(\frac{3D}{4}P-
  \frac{\partial}{\partial\eta}\left[\frac{D\eta}{4}P\right]\right)\Bigg|_{\eta = 0}
     =0 \, .
\end{equation}
The solution of this equation can be obtained by virtue of Fourier
transform with respect to $\xi$ and Laplace transform with respect
to $\eta$. The appearing equation is then solved by the method of
characteristics. Omitting the details, the solution is found to be
\begin{equation}
 P(\eta,\xi|z)=
 \frac{1}{2\pi}\intop_{-\infty}^{\infty}P_{k}(\eta|z')e^{-\rmi k \xi}\rmd
 k \, ,
  \label{p-a-f}
\end{equation}
where the Fourier component is written as
\begin{equation}
\fl
 P_{k}(\eta|z')=
 \frac{\eta}{\eta_0} \,
  \frac{\alpha_k}{\sinh(\alpha_{k}z')} \,
   \exp[-\alpha_{k} ( \eta  + \eta_{0}) \coth (\alpha_{k}z' )] \,
   \mathrm{  I}_{2}\left(\frac{ 2\,\alpha_{k}\sqrt{\eta\,\eta_{0}} }{ \sinh(\alpha_{k}z' )
    }\right).
\end{equation}
Here $\mathrm{  I}_{2}(\ldots)$ is the second-order modified
Bessel function, $\alpha_{k}=k/\sqrt{3}$ and $z'=Dz/4$.
Integrating Eq.(\ref{p-a-f}) over $\xi$ we obtain the marginal PDF
$P(\eta| \, z)$:
\begin{equation}
 P(\eta|z)=
 \frac{1}{z'}
  \frac{\eta}{\eta_0}
   \exp\left\{-\frac{\eta +\eta_{0}}{z'}\right\}
    \mathrm{  I}_{2}\left(\frac{2\sqrt{\eta \,\eta_{0}} }{z'}\right).
\end{equation}
It is easy to find that the asymptotic of the PDF for the
amplitude jitter is
\begin{equation}
 P(\eta|z) \approx \frac{ {\eta }^{3/4} }
 {2{\sqrt{\pi z'}}{ \eta_{0} }^{ 5/4 } }
  \exp\left\{ -\frac{{ ( {\sqrt{\eta }} -
            \sqrt{ \eta_{0} } ) }^2}{z'} \right\},
             \quad \eta\rightarrow\infty \, .
             \label{asim p_A}
\end{equation}
The higher-order momenta of amplitude and frequency are:
\begin{eqnarray} \label{amp-mom}
\langle \eta^n \rangle=z'^{n} \, n! \,
\mathrm{L}_{n}^{2}(-\eta_{0}/z'),
\\ \fl
 \langle\xi^{2n}\rangle=(-1)^{n}\frac{\partial^{2n}}{\partial
k^{2n}}
   \left\{
   \mathrm{sech}^{3}(\alpha_{k}z')
   \exp[-\alpha_{k}\eta_{0}\mathrm{tanh}(\alpha_{k}z')]
   \right\}\bigg|_{k=0}, \quad n=1,2, \, \, \mathrm{etc.}
\end{eqnarray}
The odd momenta of frequency vanish. In equation \eref{amp-mom}
for the amplitude momenta $\mathrm{L}_{n}^{2}(\ldots)$ are the
generalized Laguerre functions. In particular, one can obtain the
growth of the average amplitude with the propagation distance:
\begin{equation}
\langle \eta \rangle=\eta_{0}+\frac{3D}{4}\,z.
\end{equation}

\subsection{Marginal PDF for polarization angle and amplitude. WKB-solution}
Marginal PDF $P(\eta,\beta|z)$ satisfies the following FPE:
\begin{equation}
\fl \dzp=
 -\frac{\partial}{\partial\eta}\left[\frac{3D}{4}P\right]-
  \frac{\partial}{\partial\beta}\left[\frac{D\cot(2\beta)}{8\eta}P\right]+
  \frac{\partial^2}{\partial\eta^2}\left[\frac{D\eta}{4}P\right]
  +
  \frac{\partial^2}{\partial\beta^2}\left[\frac{D}{16\eta}P\right].\label{FPE-eta-beta}
\end{equation}
We will use the WKB method to obtain a solution of the above
equation \cite{JOSAB,Gaspard}. Assuming that noise amplitude, $D$,
is small, we seek the solution in the form
\begin{equation}
P=\exp[-S/D], \quad S=W+DW_1+D^2 W_2 + \ldots.
\end{equation}
In the leading order in $D$ one obtains a Hamiltonian-Jacobi
equation for the function $W$:
\begin{equation}
\frac{\partial W}{\partial z}+
 \frac{\eta}{4}\left[\frac{\partial W}{\partial \eta}\right]^{2}+
  \frac{1}{16\eta}\left[\frac{\partial W}{\partial
  \beta}\right]^{2}=0
   \label{hj}.
\end{equation}
The attached Freidlin-Wentzell Hamiltonian \cite{Gaspard} is
\begin{equation}
H=
 \frac{\eta}{4}p_{\eta}^{2}+
  \frac{1}{16\eta}p_{\beta}^{2} \, ,
\end{equation}
where $p_{\eta}$ and $p_{\beta}$ represent the ''momenta''
canonically conjugated to the state variables, $\eta$ and $\beta$
correspondingly:
\begin{equation}
p_{\eta} = \frac{\partial W}{\partial \eta} \, , \qquad p_{\beta}
= \frac{\partial W}{\partial \beta} \, .
\end{equation}
The solution of (\ref{hj}) is given by
\begin{equation}
W=\intop_{0}^{z}\left[ p_{\eta}(z')\frac{\rmd \eta (z')}{\rmd z'}+
                 p_{\beta}(z')\frac{\rmd \beta(z')}{\rmd z'} \right]\rmd z'-
                 Ez
                  \label{w} ,
\end{equation}
where $E=H$ is an integral of motion (see Refs.
\cite{JOSAB,Gaspard} for details). The Hamiltonian trajectories in
the integrand of (\ref{w}) are found from Hamilton's equations:
\numparts
\begin{eqnarray}
\frac{\rmd \eta}{\rmd z}=\frac{\eta}{2} \, p_{\eta} \, , \label{eq-Ham1} \\
\frac{\rmd \beta}{\rmd z}=\frac{1}{8\eta} \, p_{\beta} \, , \label{eq-Ham2}\\
\frac{\rmd p_\eta}{\rmd z}=-\frac{1}{4} \,
p_{\eta}^{2}+\frac{1}{16\eta^2} \, p_{\beta}^{2}, \label{eq-Ham3} \\
\frac{\rmd p_\beta}{\rmd z} = 0 \, . \label{eq-Ham4}
\end{eqnarray} \endnumparts
These trajectories are subject to the boundary conditions:
\begin{equation}
\eta(0)=\eta_{0}, \quad \eta(z)=\eta,\quad
 \beta(0)=\beta_{0}, \quad\beta(z)=\beta.
  \label{bc}
\end{equation}
The solutions for the amplitude $\eta(z)$ and the polarization
angle $\beta(z)$ are given by:
\begin{equation}
\eta(z)=
 \frac{z^2\,p_{\beta 0}^{2} +
    4\,\eta_{ 0}^{2}\,
     {\left( 4 + z\,p_{\eta 0} \right) }^2}{64\,
    \eta_{0}} \, , \label{Ham-eta}
\end{equation}
\begin{equation}\fl
\beta(z)=\beta_{0}
 -\arctan \left(\frac{2\,\eta_{ 0}\,p_{\eta 0}}
    {p_{\beta 0} }\right)
+ \arctan \left(\frac{16\,\eta_{0}^{2}\,p_{\eta 0}  +
  z\,\left( p_{\beta 0}^{2} +
     4\,\eta_{ 0}^{2}\,p_{\eta 0}^{2} \right)}{8\,
     \eta_{0}\,p_{\beta 0} }\right). \label{Ham-beta}
\end{equation}
One can now find constants $p_{\beta 0}$ and $p_{\eta 0}$:
\[
{p_{\beta 0} =
   \pm {\frac{8\,{\sqrt{\eta}}\,{\sqrt{\eta_{ 0}}}\,
       \tan (\beta - \beta_{0}) |\cos (\beta - \beta_{0})|}{z}}},
\]
\[ {p_{\eta 0}=
   {\frac{4\,\left( -{\sqrt{\eta_{ 0} }} \pm
         {\sqrt{\eta}}\,|\cos (\beta - \beta_{ 0})|
         \right) }{z\,\sqrt{\eta_{ 0}}}}} \, .
\]
From (\ref{w})--(\ref{eq-Ham4}) it follows that $W=Ez$ so that we
have two solutions of Hamilton-Jacobi equation (\ref{hj}):
\begin{equation}
W_{\pm}=\frac{4\,\left( \eta  + \eta_{0} \pm
      2\,\sqrt{\eta }\,\sqrt{\eta_{0}}\,
       |\cos (\beta  - \beta_{0})| \right) }{z} \, ,
\end{equation}
The WKB approximation for the solution of (\ref{FPE-eta-beta}) is
written as a sum
$P(\eta,\beta|z)=P_+(\eta,\beta|z)+P_{-}(\eta,\beta|z)$, where
$P_{\pm}(\eta,\beta|z)$ is given by
\begin{eqnarray}
P_{\pm}= &C \left| \det \left(-\frac{\partial^2 W_\pm}{\partial q
\partial q_0}\right) \right|^{1/2}\,\exp\left[\intop_0^z B \, p\, \rmd z'
\right] \times \nonumber \\ &\exp\left[-\frac{1}{2}\,
\intop^z_0\,\Tr \frac{\partial^2 H}{\partial q \partial p} \, \rmd
z' \right]\,\exp\left[-\frac{W_\pm}{D}\right] . \label{P_WKB}
\end{eqnarray}
Here for the sake of compactness we introduced the following 2D
vectors: $q=(\eta,\beta)$, $q_0=(\eta_0,\beta_0)$,
$p=(p_\eta,p_\beta)$. Vector $B$ represents the part of the
advection vector which is proportional to $D$ (with factor $D$
omitted). In our case $B=\{3/4,\cot (2\beta)/8\eta\}$. The
integrals in the exponential are taken along the corresponding
Hamiltonian trajectories (\ref{Ham-eta}),(\ref{Ham-beta}), with
the corresponding choice of sign in $p_{\eta 0}$, $p_{\beta 0}$.
Factor $C$ is added to provide the normalization of the PDF
$P(\eta,\beta|z)$. Each multiplier in (\ref{P_WKB}) can be
evaluated separately for both trajectories:
\[
\det \left(-\frac{\partial^2 W_\pm}{\partial q
\partial q_0}\right) = \frac{16}{z^2} \, , \]
\[ -\frac{1}{2} \,
\intop^z_0\,\Tr \frac{\partial^2 H}{\partial q
\partial p} \, \rmd z'= -\frac{1}{4}\, \intop^z_0 p_\eta \rmd z'= -\frac{1}{2}\, \ln
\left(\frac{\eta}{\eta_0}\right),
\]
and
\begin{eqnarray} \nonumber
\intop^z_0 B p\,\rmd z' &= \intop^z_0 \left(\frac{3}{4} p_\eta +
\frac{p_\beta}{8\eta} \cot(2\beta)\right) \rmd z' =
\intop^z_0\left( \frac{3}{4}\,\frac{2\eta'}{\eta}+\cot(2\beta)
\beta' \right) \rmd z'
\\ \nonumber &=
\frac{3}{2}\,\ln \left(\frac{\eta}{\eta_0}\right)+\frac{1}{2}
\ln\left|\frac{\sin(2\beta)}{\sin(2\beta_0)}\right| .
\end{eqnarray}
This yields
\[ \fl
P_\pm(\eta,\beta|\,z)=\frac{4C}{z}\, \frac{\eta}{\eta_0}\,
\sqrt{\frac{|\sin(2\beta)|}{|\sin(2\beta_0)|}}\,
\exp\left[-\frac{4}{Dz}\,\left(\eta+\eta_0 \pm 2 \sqrt{\eta\eta_0}
\, |\cos(\beta-\beta_0)|\right)\right].
\]
To obtain the normalization constant $C$ we must integrate $P_+$
and $P_{-}$ over $\eta$ and $\beta$ and insure that the sum of the
two integrals evaluates to unity. We perform the integration over
$\beta$ first. Since in the WKB approximation the noise intensity,
$D$, is assumed to be small, the main contribution into the
integral over $\beta$ comes from the vicinity of the points where
either $\cos(\beta-\beta_0)$ or $\sin(\beta-\beta_0)$ vanish. This
implies $|\sin(2\beta)| \approx |\sin(2\beta_0)|$, and one obtains
\begin{equation}
\intop^{2\pi}_{0} P(\eta,\beta|\,z) \rmd \beta \approx \frac{16\pi
C}{z}\,\frac{\eta}{\eta_0}\, \exp \left[-\frac{4}{Dz} \,
(\eta+\eta_0)\right]\,
\mathrm{I}_0\left(\frac{8\sqrt{\eta\eta_0}}{Dz}\right),
\end{equation}
where we have used the identity $\int_{0}^{2\pi} \cosh\{x |\cos
t|\} dt = 2\pi \mathrm{I}_0(x)$. The expression above is nothing
else but a WKB approximation for the marginal PDF for the
amplitude, $P(\eta|\,z)$. Since the WKB approximation is valid
under the assumption that $D \ll 1$ (or, more precisely, $Dz \ll
1$) one can use the asymptotic of the Bessel function for the
large values of its argument and observe an excellent agreement
with exact formula (\ref{asim p_A}) obtained earlier. Another
integration over $\eta$ allows one to obtain the normalization
constant $C$. The final result for the marginal PDF
$P(\eta,\beta|\,z)$ reads:
\begin{equation}\fl
P(\eta,\beta|z)\approx \frac{2}{\pi D z}\, \frac{\eta}{\eta_0}\,
\sqrt{\frac{|\sin(2\beta)|}{|\sin(2\beta_0)|}}\,\exp
\left[-\frac{4}{Dz} \, (\eta+\eta_0)\right] \cosh
\left[\frac{8\,\sqrt{\eta \,
\eta_0}}{Dz}\,|\cos(\beta-\beta_0)|\right]. \label{P-a-beta}
\end{equation}
Note that this result is inapplicable for $\sin(2\beta_0)=0$. This
is a direct consequence of the fact that the adiabatic
perturbation theory fails for such values of polarization angle
(see discussion at the end of Section \ref{sec:pertub}).

\section{Additive noise approximation}
\label{sec:Gaussian}%

Let us consider the case when the propagation distance $z$ is
small. We assume that the deviations of soliton parameters from
their initial values are small as well. Then for generic set of
Langevin equations (\ref{general}) we may linearize the advection
vector around the initial position $q_0$ and assume that the
perturbation functions $\mathbf{g}_i(q,t) $ depends on the initial
value of vector $q_0$ rather than instantaneous value $q$ (as in
Appendix we omit vector notations for $q$ and $f$). Under such
assumptions Eq.(\ref{general}) is transformed into a simple linear
Langevin system with additive Gaussian noise:
\begin{equation}
\frac{\rmd q_{i}(z)}{\rmd z}=
 f_{i}({q}_0)+\frac{\partial
 f_{i}}{\partial {q}}\bigg|_{{q}_0}({q}-{q}_{0})
 + s_{i}(z),
\end{equation}
with
\begin{eqnarray}
\langle s_{i}(z)s_{j}(z')\rangle =n_{ij}\delta(z-z'),
 \\
 n_{ij}  =2G_{ij}(q_0)=\frac{1}{2}\,\rmr\intop_{-\infty}^{\infty}
  \mathbf{g}_{i}^{*}\hat{D}\mathbf{g}_{j}\,\rmd t .
\end{eqnarray}
(The elements of the matrix $\hat{G}$ are derived in Appendix, see
(\ref{G-matrix})). The statistics of soliton parameters now become
Gaussian. Similar approach is widely used to obtain approximate
expressions for the variances of the soliton parameters in scalar
noise-driven NLSE \cite{Mecozzi}. Without loss of generality one
can assume that $q_0$ corresponds to the steady state of the
unperturbed system, i.e. $f_i(q_0)=0$. The mean and variances of
vector $\delta q=q-q_0$ are now readily obtained in a tensor form:
\begin{eqnarray}
<\delta q(z)>=0 \, ,\\
<\delta q(z)  \otimes  \delta q(z)>= 2 \intop_{0}^{z}\rmd z'
\exp[\hat{A} z'] \hat{G}(q_0)  \exp[\hat{A}^{T}z'] \, ,\\
\hat{A}=\frac{\partial f}{\partial {q}}\bigg|_{{q}_0} \, .
\end{eqnarray}
Now assuming that $\Omega_0=T_0=0$, and redefining phase $\alpha
\to \alpha - 4 \eta_0^2 z$ we obtain the following expressions for
the soliton jitters:
\begin{eqnarray}
<\delta \eta^2>=\frac{z\eta_0}{2}\,
\tilde{D}(\beta_0,\varphi_0),\label{jit-eta} \\
<\delta T^2> = \frac{z(3\pi^2+256z^2\eta_0^4)}{288\eta_0^3}\,
\tilde{D}(\beta_0,\varphi_0) \label{jit-T}, \\
<\delta \xi^2> = \frac{Z\eta_0}{6}\,\tilde{D}(\beta_0,\varphi_0),
\label{jit-xi}\\
<\delta \varphi^2>= \frac{z}{8\eta_0
\sin^2(2\beta_0)}\,\tilde{D}(\pi/2+\beta_0,\varphi_0),
\label{jit-phi} \\
<\delta
\beta^2>=\frac{1}{8\eta_0}\,\tilde{D}(\pi/2+\beta_0,\varphi_0),
\label{jit-beta}\end{eqnarray}
and
\begin{eqnarray}\nonumber
<\delta \alpha^2> &= 2\,F (\beta_0,\varphi_0) \,z + 2D_{12}\,\cot
(2\,\beta_0 )\, \sin (2\,\varphi_0 )\,z^2\, \eta
\\ &+ \frac{128}{3}\, \tilde{D}(\beta_{0},\varphi_0) \eta_0
z^{3}\,\left( \,{\eta_0 }+ \frac{\xi^2}{12} \right),
\label{jit-alpha}
\end{eqnarray}
Here we have introduced the function
\begin{equation}
\tilde{D}(\beta_0,\varphi_0)=D_{11} \cos^2\beta_0 + D_{22} \sin^2
\beta_0 + D_{12}\cos(2\varphi_0)\sin(2\beta_0),
\end{equation}
and $D_{\alpha \beta}$ are the elements of the correlation matrix,
see \eref{corrmatrix}. Formula (\ref{jit-T}) for the timing jitter
was first obtained by Doktorov and Kuten \cite{doktorov1}. The
expression for $F(\beta_0,\varphi_0)$ in \eref{jit-alpha} is
rather involved:
\begin{eqnarray}\nonumber
\fl
 F(\beta_0,&\varphi_0) = \frac{\pi^{2}\xi^{2}}{48\,\eta_{0}^{3}}
 \tilde{D}(\beta_0,\varphi_0)
  +
 \frac{1}{576\eta_0} \left\{ D_{11}
   \left[ 129 + 4\,{\pi }^2 \right]
     \,{\cos^2 (\beta_0 )} -
  36\,\left( 2\,
      \left[ D_{11} +
        D_{22} \right]  + \right.\right.
\\ \nonumber
\fl &\left.
      D_{11} \,\cos (2\,\beta_0 )
     \right)+
  9\,\cos (2\,\beta_0 )\,\left( D_{22}
      \left[ 11 +
        {\cot^2 (\beta_0 )} \right]  +
     4\,D_{12}\,
      \cos (2\,\varphi_0 )\,
      \cot (2\,\beta_0 ) \right)+
\\ \nonumber \fl
  &54 \, D_{11}\, \sin^2 (\beta_0 )\,
     +
     4\, D_{22}
      \left[ 48 + {\pi^2} \right]\sin^{2}(\beta_0)  +
      9\,{D_{11}}\,
      {\tan^2 (\beta_0 )}\sin^2 (\beta_0 )
       +
\\ \nonumber \fl
  &\left. 4\,D_{12}
   \left[ 12 + {\pi^2} \right]
   \cos (2\,\varphi_0 )\,
   \sin (2\,\beta_0 )
    \right\}.
\end{eqnarray}

Eqs.(\ref{jit-eta})--(\ref{jit-xi}) can be endowed with a simple
interpretation. First one notices that a single soliton ansatz
(\ref{u}),(\ref{v}) can be transformed into a single soliton
anzatz of the scalar NLSE via a unitary transformation:
\begin{eqnarray}
\bPsi =\hat{T}(z)\, \mathbf{\tilde{\bPsi}}, \qquad
\mathbf{\tilde{\bPsi}}= \left(
\begin{array}{c} u_0 \\ 0 \end{array} \right),  \\
T(z)=\left[\begin{array}{cc} \cos\beta(z) & -\sin\beta(z) \\
\sin\beta(z) & \cos \beta(z)
\end{array}\right]\left[\begin{array}{cc} \rme^{\rmi \varphi(z)} & 0
\\ 0 & \rme^{-\rmi
\varphi(z)} \end{array}\right] ,\label{unitary}
\end{eqnarray}
where $u_0=2\eta \,
\mathrm{sech}[2\eta(t-T)]\exp[\rmi\alpha-2\,\rmi(t-T)\xi]$ is a
single soliton solution of the scalar NLSE. Since in the additive
noise approximation we are interested only in small deviations of
the soliton parameters from their initial positions we may assume
$\hat{T}(z) \approx \hat{T}(0)$. We can then observe that applying
transformation $\hat{T}^{-1}$ leaves the l.h.s. of perturbed
Manakov equation (\ref{man-eq}) invariant. As for the noise in the
r.h.s. it is transformed into $\mathbf{\tilde n}=\hat{T}^{-1}
\mathbf{n}$, which is again AWGN with the correlation matrix
$\hat{\tilde D}=\hat{T}^{-1}\hat{D}\hat{T}$. Therefore the
variances of the amplitude, phase and frequency of Manakov soliton
$\bPsi_0$ in the additive noise approximation are equal to those
of the scalar NLSE soliton but with the normalized noise
intensity: $D \to (\hat{\tilde
D})_{11}=\tilde{D}(\beta_0,\varphi_0)$. The variances of soliton
jitters for scalar NLSE in this approximation are known
\cite{Mecozzi,JOSAB}. One can observe that formulae
(\ref{jit-eta})--(\ref{jit-xi}) coincide with those given in
\cite{Mecozzi} for amplitude, position and soliton frequency of a
NLSE soliton provided that the noise intensity is now phase and
polarization dependent $D=\tilde{D}(\beta_0,\varphi_0)$.

This has a few interesting consequences. First we observe that for
the isotropic noise when the correlation matrix $\hat D$ is
proportional to the identity matrix, $\hat D= D\,\hat{I}$,
amplitude, timing and frequency jitter do not depend on
polarization or phase at all and we recover exactly the result for
the scalar NLSE. In general case we must analyze the behavior of
the function $\tilde{D}(\beta_0,\varphi_0)$. One can readily see
that one of the local minima is given by:
\begin{equation}
\varphi=\frac{\pi}{2}, \quad \beta=\frac{1}{2}\,\arctan \left[
\frac{2D_{12}}{D_{22}-D_{11}} \right] .\label{minimum}
\end{equation}
The minimal value of the function $\tilde{D}(\beta_0,\varphi_0)$
is
\begin{equation}
\tilde{D}_{min}=\frac{D_{11}+D_{22}}{2}-\frac{1}{2}\,\sqrt{4D_{12}^2+(D_{22}-D_{11})^2}.
\label{D-min}
\end{equation}
Because of matrix $\hat{D}$ being positive definite $\det \hat{D}
\geq 0$, and $\tilde{D}_{min}$ is always non-negative, as it
should be. If $\det \hat{D} = 0$, i.e. $D_{11}D_{22}=D_{12}^2$,
then one can in principle dispose of the soliton jitters
completely (in the additive noise approximation). In practice, one
can make soliton jitters almost negligible, provided that the
determinant of the correlation matrix $\hat{D}$ is close to zero.

\section{Conclusion}
To sum up, we have considered the non-Gaussian statistics of
Manakov soliton and demonstrated the dependence of the variances
of all soliton parameters on phase mismatch $\varphi_0$ and
polarization angle $\beta_0$. We derived the FPE describing the
statistics of all soliton parameters. Using this approach we were
able to work out the analytical expressions for the marginal PDFs
for soliton amplitude $\eta$, frequency $\xi$ and polarization
angle $\beta$. The statistics of the soliton frequency and
amplitude appeared very much alike for scalar and vector NLSE in
the case when vector AWGN is isotropic. This should not be
considered as a surprise since for the isotropic perturbation of
the Manakov system there exists a unitary transformation of the
Manakov soliton into a NLSE soliton (localized in a single
polarization) which leaves equation (\ref{man-eq}) invariant
\cite{LakobaKaup}. The situation however changes when one
considers cross-correlations between noises in different
polarizations. In the presence of a noisy perturbation with
cross-correlations the symmetry related to the unitary
transformation of a vector NLSE (space rotation and phase
transformation) is broken. This means that though the relation
between Manakov and NLSE solitons still exists the statistics of
the soliton components becomes polarization and phase dependent.
Controlling the initial values of the polarization and the phase
one can effectively manage the magnitude of each soliton jitter
separately (in particular timing jitter). If the determinant of
the noise correlation matrix is small one can effectively decrease
the variance of a given soliton jitter and make it negligible.
This is impossible for an isotropic noise and for a scalar NLSE.
It means that the Manakov soliton is in general more robust
against noisy perturbations than its scalar counterpart. The
correlations between different polarization noise components can
play a positive role in suppressing soliton jitters.

The results obtained in the current paper rely on a series of
assumptions. Firstly, the use of the adiabatic perturbation theory
implies that we neglect the feedback of the linear radiation on
the soliton. This is true for relatively small distances, where
soliton parameters do not experience drastic deviations over the
very short scales. Secondly, the perturbation theory developed
here is unapplicable when the field polarization is close to
linear ($\sin(2\beta)\approx 0$). Such cases should be considered
separately (see \cite{LakobaKaup} and references therein). And
finally, we have used additive noise (or Gaussian) approximation
to obtain the variances of the soliton jitters. This approximation
works fairly well for relatively small distances ($Dz \ll 1$) and
gives reliable estimates for the variances. One should not be
thinking, however, that the statistics of the tails of the PDF are
Gaussian. In fact they are not as we have shown in Section
\ref{sec:marginal}. The FPE approach developed here may come handy
when studying the statistics of the rare events forming the tails
of the soliton PDF.

\ack  We wish to thank Sergei Turitsyn for stimulating
discussions.

\appendix
\section*{Appendix. Derivation of the Fokker-Planck equation}
\setcounter{section}{1} \label{sec:appendix} The purpose of this
Appendix is the derivation of the FPE which describes the PDF of
the system with multi-component multiplicative noise. Suppose
$q^{\,s}(z)=\left( q_{1}^{s}(z), \ldots, q_{N}^{s}(z) \right)$ is
a solution of set of $N$ stochastic equations with $M$-component
multiplicative white complex gaussian noise
\begin{equation}
 \frac{\rmd q_{i}(z)}{\rmd
z}=f_{i}(q)+\rmr\intop_{-\infty}^{\infty}\rmd t\,
\mathbf{g}^{*}_i(q,t) \mathbf{n}(t,z) ,\label{general}
\end{equation}
with
\begin{eqnarray}
\mathbf{n}(t,z) =\left(\, n_1(t,z),  \ldots ,n_M(t,z)\,\right),
\\
\mathbf{g}_{i}(q,t)=\left(\, g_{i,1}(q,t), \ldots, g_{i,M}(q,t)
\,\right).
\end{eqnarray}
and
\begin{eqnarray}
\langle n_{\alpha}(z,t)\rangle = \langle
n_{\alpha}(z,t)n_{\beta}(z',t') \rangle = 0,
\\
\langle n_{\alpha}(z,t)n_{\beta}^{*}(z',t') \rangle =
D_{\alpha\beta} \delta(z-z')\delta(t-t'). \label{cor-D}
\end{eqnarray}
Here and in the sequel we will omit vector notations for the
$N$-dimensional vectors $q$ and $f$ in order not to confuse them
with vectors $\mathbf{g}_i$ and $\mathbf{n}$ since they have
different dimensionality.

We are interested in equation for the PDF of the solution of
(\ref{general}), $P(q,z)$. This PDF can be expressed in the form
of
\begin{equation}
P(q,z)=\left\langle\,
 \delta(q-q^{\,s}(z))
\,\right\rangle.
\end{equation}
Differentiating the latter equation with respect to $z$ and
substituting the derivatives $\rmd q_{i}^s(z)/\rmd z$ one obtains
\begin{equation}
\fl
\frac{\partial P}{\partial z}=
-\sum_{i}\frac{\partial}{\partial q_i}[f_{i}P] -
\sum_{i}\rmr\frac{\partial}{\partial
q_i}\intop_{-\infty}^{\infty}\rmd t
 \left \{
\mathbf{g}_{i}^{\,*}(q,t)\left\langle\, \delta(q-q^{\,s}(z))
\mathbf{n}(t,z) \,\right\rangle \right\},
\end{equation}
where the average in the latter term can be evaluated by virtue of
Furutsu-Novikov formula \cite{Sancho} that in the case of vector
white complex noise $\mathbf{n}(t,z)$ is generalized to
\begin{equation}
 \left\langle
\Theta[\mathbf{n},\mathbf{n}^{\,*}]\mathbf{n}(t,z) \right\rangle=
 \hat{D}\left\langle  \frac{\delta \Theta[\mathbf{n},\mathbf{n}^{\,*}]}{\delta\mathbf{n}^{\,*}(t,z)}
 \right\rangle \, ,
\end{equation}
where $\Theta[\mathbf{n},\mathbf{n}^{\,*}]$ is an arbitrary
functional of noise and matrix $\hat D$ has elements
$D_{\alpha\beta}$ (\ref{cor-D}). Since
\begin{eqnarray}
\frac{\delta n_{\alpha}^{*}(t,z)}{\delta n_{\beta}^{*}(t',z')}&=&
 \delta_{\alpha\beta}\,\delta(z-z')\delta(t-t') \, , \\
\frac{\delta n_{\alpha}(t,z)}{\delta n_{\beta}^{*}(t',z')}&=&0 \, , \\
\frac{\delta q_{k}^{s}(z)}{\delta \mathbf{n}^{\,*}(t,z)} &=&
\frac{1}{2}\,\theta(0)\,\mathbf{g}_{k}(q^{\,s}(z),t) \, , \qquad
\theta(0)=\frac{1}{2} \, ,
\end{eqnarray}
(where in the latter formula we have used Eq.(\ref{general}) and
assumed symmetric Stratonovich regularization of the white noise)
we arrive at
\begin{equation}
\langle \delta(q-q^{\,s}(z))\,\mathbf{n}(t,z)\, \rangle=
 -\frac{1}{4}
   \sum_{k}\frac{\partial}{\partial
   q_{k}}[\hat{D}\mathbf{g}_{k}(q,t)P(\mathbf{q},z)].
\end{equation}
Therefore for the PDF $P(q,z)$ we arrive at the FPE equation:
\begin{equation}
\frac{\partial}{\partial z}P( {q},z)=
 -\sum_{i}\frac{\partial}{\partial q_i}[G_{i}(q)P(q,z)]
 +\sum_{i,k}\frac{\partial^{2}}{\partial q_{i}\partial
  q_{k}}[G_{ik}(q)P(q,z)].
\end{equation}
Here the components of the advection vector and the diffusion
matrix ($G_i$ and $G_{ij}$ respectively) are defined as
\begin{eqnarray}
G_{i}(q)&=&
 f_{i}(q) + \frac{1}{4}\rmr\sum_{k}\intop_{-\infty}^{\infty}\rmd t
 \,
   \frac{\partial \mathbf{g}_{i}^{\,*}}{\partial
  q_k} \hat{D}\mathbf{g}_{k},
\\
G_{ik}(q)&=&\frac{1}{4}\rmr\intop_{-\infty}^{\infty}\rmd t \,
\mathbf{g}_{i}^{\,*}\hat{D}\mathbf{g}_{k}, \label{G-matrix}
\end{eqnarray}
where $i,k=1, \ldots N$.

\end{document}